\documentclass[conference]{IEEEtran}
\IEEEoverridecommandlockouts

\usepackage{cite}
\usepackage{amsmath,amssymb,amsfonts}
\usepackage{algorithmic}
\usepackage{graphicx}
\usepackage{textcomp}
\usepackage{xcolor}
\usepackage{comment}
\usepackage{enumitem}
\usepackage{ulem}
\usepackage{booktabs}
\usepackage{tabularx}
\usepackage{footnote}
\usepackage{soul}
\usepackage{url}
\usepackage{hyphenat}
\usepackage{balance}
\usepackage[listings]{tcolorbox}

\def\BibTeX{{\rm B\kern-.05em{\sc i\kern-.025em b}\kern-.08em
    T\kern-.1667em\lower.7ex\hbox{E}\kern-.125emX}}
\begin{document}


\title{\textsc{CIAO} - \textsc{Code In Architecture Out} - Automated Software Architecture Documentation with Large Language Models}

\author{\IEEEauthorblockN{Marco De Luca}
\IEEEauthorblockA{\textit{University of Naples Federico II}\\
Naples, Italy \\
marco.deluca2@unina.it}
\and
\IEEEauthorblockN{Tiziano Santilli}
\IEEEauthorblockA{\textit{University of Southern Denmark}\\
Odense, Denmark \\
tisa@mmmi.sdu.dk}
\and
\IEEEauthorblockN{Domenico Amalfitano}
\IEEEauthorblockA{\textit{University of Naples Federico II}\\
Naples, Italy \\
domenico.amalfitano@unina.it}

\and
\IEEEauthorblockN{Anna Rita Fasolino}
\IEEEauthorblockA{\textit{University of Naples Federico II}\\
Naples, Italy \\
fasolino@unina.it}

\and
\IEEEauthorblockN{Patrizio Pelliccione}
\IEEEauthorblockA{\textit{Gran Sasso Science Institute} \\
L’Aquila, Italy\\
patrizio.pelliccione@gssi.it}
}

\maketitle

\begin{abstract}
Software architecture documentation is essential for system comprehension, yet it is often unavailable or incomplete. While recent LLM-based techniques can generate documentation from code, they typically address local artifacts rather than producing coherent, system-level architectural descriptions. This paper presents a structured process for automatically generating system-level architectural documentation directly from GitHub repositories using Large Language Models. The process, called CIAO (Code In Architecture Out), defines an LLM-based workflow that takes a repository as input and produces system-level architectural documentation following a template derived from ISO/IEC/IEEE 42010, SEI \textit{Views \& Beyond}, and the C4 model. The resulting documentation can be directly added to the target repository.
We evaluated the process through a study with 22 developers, each reviewing the documentation generated for a repository they had contributed to. The evaluation shows that developers generally perceive the produced documentation as valuable, comprehensible, and broadly accurate with respect to the source code, while also highlighting limitations in diagram quality, high-level context modeling, and deployment views. We also assessed the operational cost of the process, finding that generating a complete architectural document requires only a few minutes and is inexpensive to run.
Overall, the results indicate that a structured, standards-oriented approach can effectively guide LLMs in producing system-level architectural documentation that is both usable and cost-effective.
\end{abstract}


\section{Introduction}

Software architecture documentation plays a central role in supporting system comprehension, communication, and long-term evolution. When documentation is missing, outdated, or inconsistent with the implementation, developers struggle to understand system decomposition, responsibilities, and dependencies, often leading to architectural drift, erosion, and ultimately architectural technical debt~\cite{Garlan09,Lago20,Kruchten2012,Minh2018}. Despite its recognized importance, architectural documentation is frequently incomplete or informal in industrial and open-source projects, where time pressure and code-centric practices make its production difficult to sustain. Established standards provide guidance on how architectural information should be documented. ISO/IEC/IEEE 42010 defines core concepts and the relationship between architecture descriptions and stakeholder concerns, emphasizing that documentation should explicitly address the needs of its intended readers~\cite{ISO42010}. SEI's \textit{Views \& Beyond} framework promotes a view-based, stakeholder-oriented approach in which different views capture complementary architectural structures~\cite{Clements2010}. In line with these principles, we focus on \textbf{system-level} architectural documentation aimed at developers who require a consolidated, high-level understanding of a repository.

Recent years have seen the rapid adoption of Large Language Models (LLMs) in software engineering, with growing evidence of their usefulness in tasks such as program comprehension, code summarization, documentation generation, and automated repair~\cite{Fan2024Survey,Zhang2023Survey,llm5,llm6}. Industrial analyses, such as GitHub’s \textit{Octoverse} report, similarly highlight the increasing reliance on LLM-based assistants to help developers navigate large codebases~\cite{GitHubOctoverse}. Research has explored LLMs for generating fine-grained documentation artifacts, such as API descriptions~\cite{Chaplia24}, code explanations~\cite{Nam24}, test-case summaries~\cite{Djajadi25}, and requirement-like statements extracted from source code~\cite{Xu25}. Other works explored LLMs for supporting higher-level reasoning, such as extracting UML diagrams, identifying design patterns, or reconstructing domain models from code~\cite{Siala25,Miranda24,Boronat25,Chen23}. 
These contributions highlight the potential of LLMs to support documentation and architecture recovery tasks. However, most approaches focus on specific artifacts or narrow tasks. 
The use of LLMs to produce system-level architectural documentation directly from repositories, following established documentation standards, remains largely unexplored.

This paper addresses this gap by introducing \textsc{CIAO} (Code In Architecture Out), a structured process for automatically generating system-level architectural documentation from  GitHub repositories using LLMs. CIAO defines a workflow that takes a repository as input and produces an architectural documentation following a template, which has been defined by taking inspiration from ISO/IEC/IEEE 42010, SEI \textit{Views \& Beyond}, and the C4 model~\cite{BrownC4}. The final output is an architectural documentation, which can be directly integrated into the target repository as a README file. 
Our study evaluates CIAO with 22 developers who analyzed the documentation generated for repositories they contributed to. The questionnaire addresses perceived value (RQ1), comprehensibility (RQ2), accuracy (RQ3), and identified limitations (RQ4). We also measure the generation time and computational cost (RQ5).

The main contributions of this work are as follows:
\begin{itemize}
    \item A standards-oriented template for system-level architectural documentation, based on ISO/IEC/IEEE~42010, SEI \textit{Views \& Beyond}, and the C4 model.
    \item A structured LLM-based workflow that generates architectural documentation directly from GitHub repositories.
    \item An open-source prototype implementing the proposed workflow, capable of producing ready-to-use architectural documentation that can be directly integrated into the target repository.
\end{itemize}

The remainder of this paper is organized as follows.
Section~\ref{sec:llm_doc_relwork} reviews related work. Section~\ref{sec:doc_process} presents our documentation process. Section~\ref{sec:evaluation} describes the experimental evaluation, and Section~\ref{sec:Results} reports the results.  Section~\ref{sec:threats} discusses the threats to validity. Finally, Section~\ref{sec:conclusion} concludes the paper and outlines future work.

The supplementary material, which includes the CIAO implementation, the generated documentation, and the full survey structure and results, is available at the following link: \url{https://doi.org/10.5281/zenodo.18710540}

\section{Related Work}
\label{sec:llm_doc_relwork}

Reverse engineering comprises techniques for reconstructing the structure, behavior, and design intent of software systems from low-level artifacts such as source code and execution traces~\cite{Nelson2005}. It is particularly valuable when documentation is missing or outdated, helping developers regain understanding of system decomposition, component relationships, and design decisions~\cite{Stringfellow06}. A central branch of this field is Software Architecture Recovery (SAR), which focuses on rebuilding the architectural organization of a system by identifying its main components, dependencies, and structural patterns~\cite{Telea02,Garcia13}. SAR is particularly relevant when architectural drift or erosion causes the implemented architecture to diverge from the intended one~\cite{Minh2018,DESILVA2012132,Garlan09,Lago20,bucaioni2025architecture,Wohlrab2019}. When these inconsistencies are not addressed in architectural descriptions, they become architectural debt~\cite{Kruchten2012}. Existing SAR approaches include static analysis, dynamic analysis, and techniques using ranking, clustering, or machine learning to infer architectural boundaries~\cite{Syst2000,Revealer,Sora16}. Across these techniques, the goal is to provide maintainers with a coherent architectural view that is otherwise implicit in large codebases.

The rapid adoption of LLMs in software engineering has led to multiple documentation-related approaches that vary in input artifacts, abstraction level, and produced outputs. LLMs have been used for summarization and automated repair~\cite{llm5,llm6,Ahmed24,Zhang24,nejjar2023llms}, with prompt-driven interaction enabling flexible, natural-language guidance~\cite{Liu23}.
A first group of studies uses source code as input to derive natural-language documentation or higher-level descriptions. Applications include extracting REST API endpoints for microservice systems~\cite{Chaplia24}, generating context-aware explanations during program comprehension~\cite{Nam24}, and producing requirement-like statements through the \textsc{AutoReqGen} pipeline~\cite{Xu25}. Other contributions focus on code fragments, such as intent-oriented summaries~\cite{Nugroho25} or explanations of legacy languages~\cite{Diggs25}, and on test artifacts, where LLMs generate concise summaries to support navigation~\cite{Djajadi25}. Recent studies have investigated whether LLMs can support higher-level reasoning, such as identifying design patterns, generating UML diagrams, or reconstructing domain models~\cite{Miranda24,Chen23,Amalfitano26}. Hybrid approaches combine static analysis with LLM reasoning to recover structural views~\cite{Siala25}, while other studies compare LLM-based extraction with traditional MDRE techniques~\cite{KevinLano}. \textsc{MDRE-LLM} leverages RAG techniques to recover domain models at different granularities~\cite{Boronat25}. Additional work has explored architectural reasoning from requirements or code in microservice settings~\cite{Quevedo24,Gustrowsky24,Jahic24}. LLMs have also been used to generate textual descriptions from UML Use Case Diagrams~\cite{Naimi24} or to support modeling tasks in educational contexts~\cite{Garaccione}. A third line of work explores how LLMs can enrich or review existing documentation. Examples include enhancing API documentation with usage examples~\cite{Naghshzan25}, tailoring documentation to different stakeholders~\cite{Bala25}, and supporting documentation review processes~\cite{Demir25}.

A recent contribution closely related to our work is \textit{CodeDocs-GenAI}~\cite{CodeDocs}, which uses RAG and LLM-based summarization to generate README documentation for Git repositories. Like CIAO, it analyses the entire repository, including its structure and source code, to generate structured documentation. However, its focus remains on general-purpose project overviews rather than system-level architectural documentation grounded in established standards and guidelines.

Overall, prior work shows that LLMs have been applied to many documentation-related tasks, mostly targeting specific artifacts such as APIs, code snippets, tests, requirements, or individual diagrams. By contrast, their use for full architecture recovery and end-to-end architectural documentation directly from source code remains largely unexplored. Our work addresses this gap by operating at a higher level of granularity: starting from a repository, we generate comprehensive architectural documentation across multiple views, stakeholders, and concerns within a unified template, extending the role of LLMs from local documentation support to system-level architectural documentation.

\section{The Documentation Process}
\label{sec:doc_process}

In this section, we present the automated documentation process implemented in \textsc{CIAO}, which generates \textit{system-level architectural documentation} from GitHub repositories using an LLM. The process is built around a system-level architectural template that is lightweight enough to be provided as input to the LLM while remaining consistent with established standards and practitioner guidelines.
Section~\ref{subsec:proposed-template} introduces the template and its expert validation, Section~\ref{subsec:workflow} describes the overall workflow, Section~\ref{subsec:prompt-engineering} details the prompt-engineering strategy, and Section~\ref{subsec:model-selection} discusses the selection of the underlying LLM.

\subsection{The Proposed Template}
\label{subsec:proposed-template}

Software architecture documentation is commonly described as \textit{view-based}~\cite{book_sei}: the information to be presented and the appropriate level of abstraction should depend on stakeholders’ concerns and the architectural perspectives to be supported. In this work, we focus on \textit{system-level architectural documentation} aimed at developers who need a consolidated, high-level understanding of a repository to work with and evolve the system. The template must therefore remain aligned with established architectural practices while being lightweight and regular enough to be used as input to an LLM.

\noindent \textbf{Template Design.}
To determine which architectural information should be included and how it should be organized and presented, we adopted an expert-driven iterative design process.
Three experts participated in this phase: one software engineering researcher with ten years of experience in architecture and documentation, and two industry software architects with five years of experience each. Through three separate focus groups, we elicited the architectural elements they deemed essential at system level. The sessions converged on a common set of needs: a clear definition of system scope and external context, a representation of deployable units and their responsibilities, an internal structural view of modules and their relationships, explicit links to code artifacts, and attention to cross-cutting concerns and quality-related rationale. The resulting synthesis was consolidated by two authors and validated by a third.

These needs were then organized into a coherent structure drawing on three well-established sources. 
ISO/IEC/IEEE~42010~\cite{ISO42010_2022} guided the inclusion of concepts such as system scope, stakeholder concerns, 
external interactions, and architectural rationale. 
SEI’s \textit{Views \& Beyond}~\cite{book_sei} informed the separation between structural, behavioral, 
and deployment perspectives. 
The C4 model \cite{C4Model}, aligned with the Architecture-as-Code paradigm~\cite{bucaioni2025architecture}, 
provided a four-level hierarchy, \textit{Context} (L1), \textit{Container} (L2), \textit{Component} (L3), and \textit{Code} (L4), that is widely used for structuring developer-oriented documentation~\cite{Jongeling_C4}. 
A revised draft was then discussed in a second round of focus groups, and a final joint session with all experts was used to resolve remaining ambiguities and validate the structure.

The resulting template comprises eight sections, each addressing a distinct architectural concern and drawing consistently on ISO/IEC/IEEE~42010, SEI’s \textit{Views \& Beyond}, and the four abstraction levels of the C4 model (Context, Container, Component, Code):


\begin{enumerate}

  \item \textbf{System Overview.}
  Provides an entry point to the architectural description by summarizing the system’s purpose, scope, and main responsibilities. It establishes the \textit{system-of-interest}, in line with ISO~42010, and offers a conceptual anchor for the architectural views that follow, helping readers quickly situate the repository before moving to more technical details.

  \item \textbf{Architectural Context.}
  Characterizes the system’s external environment by identifying actors, interacting systems, APIs, and data sources. By clarifying boundaries and external dependencies, this section supports ISO’s emphasis on contextualizing the architecture and corresponds to the C4 Level 1 Context view. It provides the background necessary to understand integration points and the role of the system within a broader ecosystem.

  \item \textbf{Containers.}
  Describes the system’s logical runtime organization, following the notion of “containers’’ in the C4 model, namely, the applications and data stores that must be running for the system to operate. For each container, the documentation summarizes responsibilities, exposed interfaces, key technologies, and interaction patterns. This section offers a coarse-grained view of how major building blocks collaborate at runtime, aligning with SEI’s Component-and-Connector perspective and ISO’s runtime-oriented concerns.

  \item \textbf{Components.}
  Presents the internal logical structure of the system by identifying key modules, packages, or classes and the structural relationships among them. This section provides a technology-independent representation of architectural organization, complementing the container perspective by revealing how domain responsibilities are grouped and how subsystems interact. It aligns with SEI’s Module viewtype and corresponds to C4 Level~3.

  \item \textbf{Code-Level.}
  Connects architectural abstractions to their concrete implementation by mapping components to source-code artifacts. It identifies relevant directories, files, entry points, and recurring design or architectural patterns. By making the realization of architectural elements explicit, this section supports traceability to implementation as recommended by ISO~42010 and corresponds to C4 Level~4.

  \item \textbf{Cross-Cutting Concerns.}
  Summarizes concerns that influence multiple parts of the system, such as security, configuration, logging, testing, and monitoring, and describes how they manifest in the codebase. These aspects are critical to understanding system behavior beyond structural decomposition and align with SEI’s “beyond-the-views’’ guidance and ISO’s focus on capturing stakeholder concerns that transcend individual elements.

  \item \textbf{Quality Attributes and Rationale.}
  Highlights the quality attributes (e.g., performance, maintainability, scalability, security) supported by the implementation and synthesizes the rationale inferred from observable design choices. This section implements ISO~42010’s recommendation to document architectural rationale and supports readers in understanding why certain architectural decisions were made.

  \item \textbf{Deployment.}
  Characterizes the system’s operational infrastructure by describing deployment artifacts (e.g., Dockerfiles and configuration files), execution environments, storage and compute nodes, and their relationships. It shows how software elements map to physical, virtual, or containerized resources, aligning with the SEI Allocation perspective and C4 deployment practices, and clarifies how the system runs in practice and how runtime responsibilities are distributed.

\end{enumerate}

\subsection{CIAO Workflow}
\label{subsec:workflow}

\begin{figure}[h!]
  \centering
  \includegraphics[width=0.98\linewidth]{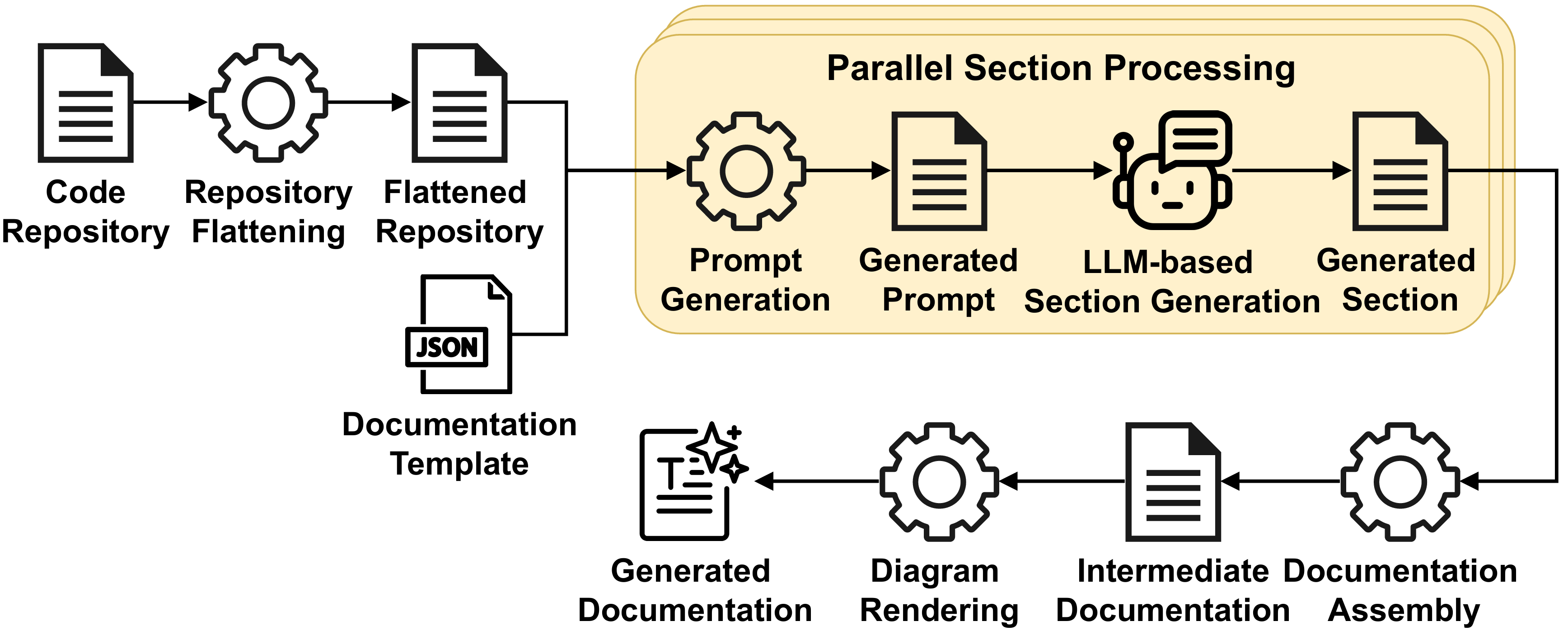}
  \caption{Overview of the CIAO automated documentation workflow.}
  \label{fig:workflow}
\end{figure}

\textsc{CIAO} is implemented as a Python-based tool that automatically generates system-level architectural documentation from GitHub repositories using LLMs. Figure~\ref{fig:workflow} summarizes the end-to-end workflow. The process takes two primary inputs: (i) the target \textit{GitHub repository}, which provides the source artifacts to be analyzed, and (ii) the \textit{documentation template}, which defines the structure of the architectural document, the goal of each section, and the global writing guidelines. Based on these inputs, \textsc{CIAO} performs four automated steps: repository flattening, prompt generation, section generation, and final assembly with diagram rendering.

\noindent \textbf{Repository Flattening:}
The repository is converted into a single textual artifact using \textsc{Repomix} \cite{repomix}, which aggregates the project’s source code into one AI-friendly file. \textsc{Repomix} can be configured to apply different filters; in our case, we remove comments and exclude files or directories that do not contain source code (e.g., binaries, build outputs, large test datasets, generated documentation), while retaining essential configuration artifacts (such as \texttt{Dockerfile}, container-orchestration descriptors, and dependency manifests like \texttt{package.json} or \texttt{pom.xml}). We also enable the option to enrich the output with a textual description of the repository structure. The resulting \textit{Flattened Repository} representation captures both the project’s folder layout and the curated code base, and serves as input for the subsequent steps.

\noindent \textbf{Prompt Generation:}
For each documentation section, \textsc{CIAO} builds, in parallel, a composite prompt consisting of two parts: (i) a fixed \textit{global prompt}, shared across all sections,  and (ii) a \textit{section-specific prompt}, derived from the \textit{Documentation Template}. 
Moreover, the \textit{Flattened Repository} representation is appended to each prompt to ensure code-grounded generation.

\noindent \textbf{LLM-based Section Generation:}
For each prompt generated in the previous step, \textsc{CIAO} submits it to the LLM, processing all sections in parallel. This step produces the set of \textit{Generated Sections}, each corresponding to a specific part of the template and adhering to its prescribed structure and constraints.

\noindent \textbf{Documentation Assembly:}
The \textit{Generated Sections} produced in the previous step are assembled into a single document, referred to as the \textit{Intermediate Documentation}. This yields a complete system-level architectural description that follows the structure defined by the template.

\noindent \textbf{Diagram Rendering:}
Since LLM-generated diagrams are provided in textual PlantUML format \cite{plantuml}, \textsc{CIAO} renders them as images and replaces the textual definitions accordingly. Applied to the \textit{Intermediate Documentation}, this step produces the final \textit{Generated Documentation}, in which all diagrams are available in visual form.

\subsection{Prompt Engineering}
\label{subsec:prompt-engineering}

\textsc{CIAO} employs a structured prompt-engineering strategy to maximize architectural accuracy, limit hallucinations, and ensure consistency across sections. Each prompt consists of (i) a \textit{global prompt}, shared across all sections, which defines the LLM’s role, target audience, writing style, and grounding requirements; and (ii) a \textit{section-specific prompt}, instantiated from the documentation template and tailored to the goal and expected artifacts of that section. This separation reflects the fact that architectural documentation comprises heterogeneous elements that require different abstraction levels and extraction strategies. This design follows the \textit{task-decomposition} prompting strategy proposed by Liu et al.~\cite{Liu2026comprehensive}, guiding the LLM to address a sequence of smaller, section-specific subtasks rather than generating the entire documentation in a single step.

\textsc{CIAO} further follows Liu et al.'s taxonomy~\cite{Liu2026comprehensive} by adopting both \textit{profile} and \textit{instruction} prompting, specifying who the model should act as and how it should perform the task.
Along the profile dimension, \textit{role prompting}~\cite{role_prompting} positions the model as a ``\textit{Meticulous Software Architect}'', helping to stabilize tone, terminology, and stylistic coherence across sections.  Additionally, light motivational cues~\cite{emotion_prompting} encourage careful reasoning. On the instruction side, prompts specify evidence-grounded generation requirements and explicitly forbid inventing architectural elements not present in the repository. Selective few-shot examples~\cite{fewshot} (e.g., small Markdown or PlantUML skeletons) provide structural scaffolding that guides formatting and level of detail. This combination supports coherent, template-aligned, and verifiable architectural documentation. The complete prompt is available in the supplementary material.

\subsection{Model Selection}
\label{subsec:model-selection}

To select the LLM used in \textsc{CIAO}, we conducted an exploratory pilot study comparing four state-of-the-art models: \texttt{GPT-5}, \texttt{Claude Sonnet 4.5}, \texttt{Gemini 2.5}, and \texttt{Mistral Large 2}. For three representative repositories provided by developers familiar with the projects, we generated documentation with each model and discussed the outputs in dedicated focus groups. The evaluation considered (i) accuracy of architectural elements and diagrams, (ii) consistency of terminology and relationships across sections, (iii) the presence of hallucinated or speculative content, and (iv) adherence to the template. Across models, \texttt{GPT-5} consistently produced the most accurate, stable, and template-aligned documentation, particularly in sections requiring multi-level structural reasoning. It also yielded more syntactically correct \textsc{PlantUML} diagrams and fewer speculative elements. Based on these observations, \texttt{GPT-5} was selected as the default model in \textsc{CIAO}, although the tool remains configurable to support alternative cost–quality trade-offs.

\section{Experimental Evaluation}
\label{sec:evaluation}

The goal of this study is to evaluate the effectiveness of the proposed LLM-based process in generating system-level architectural documentation directly from source code while adhering to our standards-oriented template. We assess the perceived value, comprehensibility, and accuracy of the generated documentation, as well as the limitations and missing aspects identified by practitioners. To this end, we conducted a survey-based evaluation in which software developers provided real-world repositories and reviewed the documentation produced by our tool. A direct comparison with related approaches was not feasible, as the only closely related work, i.e., CodeDocs-GenAI~\cite{CodeDocs}, does not provide an openly accessible implementation.
The validation aims to answer the following Research Questions (RQs):

\begin{enumerate}[label=\hspace{.2cm}RQ\arabic*:, leftmargin=.38in]
    \item \textit{Do developers perceive the documentation as valuable enough to be integrated into their own projects?}
    \item \textit{To what extent is the documentation comprehensible?}
    \item \textit{To what extent is the documentation accurate with respect to the system’s source code?}
    \item \textit{What limitations or missing aspects do developers identify in the documentation?}
    \item \textit{What are the costs of generating the documentation?}
\end{enumerate}

\begin{table*}[ht]
\centering
\scriptsize
\setlength{\tabcolsep}{3pt}
\renewcommand{\arraystretch}{0.95}
\caption{Overview of analyzed repositories with languages and code size.}
\label{tab:repo-stats}
\begin{tabularx}{\textwidth}{l l X r r}
\hline
\textbf{Repo} & \textbf{Domain} & \textbf{Lang. (LOC \%)} & \textbf{Files} & \textbf{LOC} \\
\hline\hline
Group-Key-Phemapn \cite{GroupKeyPhemap} 
  & IoT 
  & C (17.0\%), C++ (83.0\%) 
  & 9 & 1061 \\\hline

pyALS-RF-tmr \cite{pyALS_RF} 
  & Machine Learning 
  & C (0.9\%), C++ (0.3\%), CMake (1.6\%), JSON (0.7\%), Python (95.8\%), Shell (0.7\%) 
  & 85 & 9270 \\\hline

sa-bsn \cite{SA_BSN} 
  & Healthcare 
  & C (12.6\%), C++ (75.5\%), CMake (4.1\%), Python (4.5\%), Shell (0.5\%), XML (2.9\%) 
  & 175 & 9877 \\\hline

ocean-lib \cite{ocean_lib} 
  & Machine Learning  
  & Python (80\%), YAML (20\%) 
  & 33 & 1208 \\\hline

robethichor \cite{robethichor} 
  & Robotic Systems 
  & CMake (1.7\%), Dockerfile (1.6\%), JSON (3.7\%), Python (88.1\%), XML (4.9\%) 
  & 28 & 697 \\\hline

Openjob \cite{springboot_ms} 
  & Web Application 
  & Dockerfile (0.6\%), Java (65.3\%), XML (32.3\%), YAML (1.9\%) 
  & 146 & 5386 \\\hline

DSP-Frontend \cite{DSP_public} 
  & Cybersecurity 
  & CSS (10.4\%), JS (88.7\%), XML (0.9\%) 
  & 447 & 238951 \\\hline

DSP-Backend \cite{DSP_app} 
  & Cybersecurity 
  & JS (100\%) 
  & 38 & 6870 \\\hline

QuFI \cite{QuFI} 
  & Quantum computing 
  & Python (100.0\%) 
  & 8 & 851 \\\hline

BoMoDT \cite{BoMoDT} 
  & Digital Twins 
  & CSS (0.1\%), JSON (0.9\%), Python (9.7\%), XML (89.0\%), YAML (0.4\%) 
  & 55 & 50928 \\\hline

m2dt \cite{m2dt} 
  & Digital Twins
  & JSON (0.2\%), Python (86.1\%), XML (13.7\%) 
  & 36 & 4040 \\\hline

GymportalService \cite{GymportalService} 
  & Web Application 
  & Java (97.7\%), XML (2.3\%) 
  & 35 & 2458 \\\hline

pyicub \cite{pyicub} 
  & Healthcare Robotics  
  & JSON (31.2\%), Python (64.3\%), Shell (2.3\%), XML (1.2\%), YAML (0.9\%) 
  & 123 & 8441 \\\hline

treeco \cite{treeco} 
  & Machine Learning 
  & Python (100.0\%) 
  & 61 & 6959 \\\hline

TestingRobotChallenge-T4 \cite{A13} 
  & Web Application 
  & Batch (8\%), Java (71\%), Shell (11\%), XML (7\%), YAML (3\%) 
  & 48 & 1660 \\\hline

rse2025 \cite{rse2025} 
  & Web Application  
  & JS (1.2\%), YAML (98.8\%) 
  & 4 & 81 \\\hline

write2audiobook \cite{write2audiobook} 
  & Accessibility tools  
  & Python (94.1\%), YAML (5.9\%) 
  & 12 & 1080 \\\hline

BF-JITcompiled \cite{BF_JITcompiled} 
  & JIT Compiler
  & C (6.3\%), C++ (93.7\%) 
  & 3 & 668 \\\hline

visual\_novel\_pyarcade \cite{visual_novel_pyarcade} 
  & Visual novel engine  
  & JSON (7.0\%), Python (93.0\%) 
  & 18 & 1269 \\\hline

SAM-CyFra \cite{SAM_CyFra} 
  & Network Security 
  & INI (0.1\%), Makefile (0.1\%), Python (91.5\%), SQL (7.5\%), XML (0.9\%) 
  & 131 & 9796 \\\hline
  
  SCASS \cite{SCASS} 
  & Cybersecurity 
  & Dockerfile (1.8\%), JSON (33.8\%), JS (0.2\%), Makefile (0.1\%), Perl (6.9\%), Prolog (9.6\%), Python (15.1\%), Shell (1.5\%), YAML (31.1\%) 
  & 58 & 5088 \\\hline

OSINT \cite{OSINT} 
  & Cybersecurity 
  & CSS (1.0\%), Dockerfile (0.4\%), Go (3.6\%), JSON (71.6\%), Java (7.0\%), JS (0.7\%), Python (11.7\%), Shell (1.1\%), TypeScript (1.2\%), XML (0.3\%), YAML (1.4\%) 
  & 172 & 23999 \\\hline

\end{tabularx}
\end{table*}

\noindent \textbf{Metrics.}
To answer our research questions, we rely on two complementary sets of measurements. For RQ1–RQ4, we base our evaluation on the metrics derived from a questionnaire that we designed for this study. The questionnaire includes both Likert-scale items and open-ended questions, and the corresponding items for each research question are described in the following sections. For RQ5, we assess the cost of the documentation-generation process by measuring two quantitative indicators: (i) the time required to generate the complete architectural documentation for each repository, and (ii) the financial cost associated with the model’s API usage. These measurements allow us to evaluate both the technical and economic impact of the process.

\noindent \textbf{Questionnaire Design.}
To address our research questions, we constructed a questionnaire composed of closed-ended items measured on a 5-point Likert scale (\emph{1 = strongly disagree (SD)}, \emph{2 = disagree (D)}, \emph{3 = neutral (N)}, \emph{4 = agree (A)}, \emph{5 = strongly agree (SA)}) and open-ended questions designed to elicit richer qualitative insights.

The introductory section collects background information about each participant and their relationship with the evaluated repository. Specifically, it records job role, application domain (e.g., web application, automotive, healthcare), prior familiarity with the codebase, and whether the participant had previously authored or maintained any repository documentation (e.g., READMEs, architectural descriptions, wiki pages, ADRs). When applicable, participants briefly described the type of documentation they had produced.This information helped contextualize their assessments and interpret perception differences across levels of experience and prior knowledge.

The second part of the questionnaire is organized around the four research questions. Table~\ref{tab:questionnaire-structure} summarizes all items, including identifiers (Q1–Q30), wording, and type.  
Perceived value (\textbf{RQ1}) is examined through three Likert-scale items and one open-ended question (Q1–Q4), probing the usefulness and perceived contribution of the generated documentation.  
Comprehensibility (\textbf{RQ2}) is evaluated through four Likert-scale items and one open-ended question (Q5–Q9) focusing on clarity, structure, terminology, and redundancy.  
Accuracy and consistency (\textbf{RQ3}) are assessed using the largest set of items (Q10–Q27), which combine global questions on architectural correspondence, correctness of responsibilities and relationships, and internal consistency with section-level items aligned with the proposed template (e.g., Containers, Components, Code-Level, Use Case Diagram, and Deployment Diagram), followed by an open-ended prompt for additional comments.  
Limitations and missing aspects (\textbf{RQ4}) are captured through three open-ended questions (Q28–Q30) targeting structural or conceptual weaknesses, missing information, and potential improvements.
The questionnaire was implemented in Google Forms, which was used to collect all participants’ responses.

\begin{table*}[t]
\centering
\renewcommand{\arraystretch}{0.8}
\caption{Overview of the questionnaire structure, question types, and mapping to questionnaire parts. Legend: L = 5-point Likert scale; O = Open-ended question.}
\label{tab:questionnaire-structure}
\scriptsize
\begin{tabularx}{\textwidth}{p{0.2cm} X r}
\toprule
\textbf{ID} & \textbf{Question} & \textbf{Type} \\
\midrule
\multicolumn{3}{l}{\textbf{Part 1 – Perceived value (RQ1)}} \\
\midrule
Q1  & The LLM-generated architectural documentation provides valuable architectural insights about the system. & L \\
Q2  & The LLM-generated documentation is useful for understanding and maintaining the system’s structure and dependencies. & L \\
Q3  & The LLM-generated documentation is valuable enough that I would consider using it in my own project. & L \\
Q4  & Which sections or features of the LLM-generated documentation were most or least valuable to you? & O \\
\midrule
\multicolumn{3}{l}{\textbf{Part 2 – Comprehensibility (RQ2)}} \\
\midrule
Q5  & The LLM-generated architectural documentation is clear, well-structured, and easy to follow. & L \\
Q6  & The LLM-generated architectural documentation uses appropriate terminology for software architecture. & L \\
Q7  & The LLM-generated architectural documentation contains excessive redundancy or unnecessary information. & L \\
Q8  & The explanations within the documentation are sufficiently detailed. & L \\
Q9  &  Which parts of the documentation were most or least comprehensible to you? & O \\
\midrule
\multicolumn{3}{l}{\textbf{Part 3 – Accuracy and consistency (RQ3)}} \\
\midrule
Q10 & The LLM-generated documentation accurately reflects the architecture implemented in the repository. & L
\\
Q11 & The documentation is internally consistent across sections and diagrams. & L \\
Q12 & The content is complete and does not omit key architectural elements. & L \\
Q13 & The relationships and dependencies among components are well captured & L \\
Q14 & The component responsibilities are correctly described in the documentation & L \\
Q15 & I would use the documentation as a reliable reference for the system architecture. & L \\
Q16 & Section 1: System Overview: The purpose, scope, and key features described are correctly derived from the repository content. & L \\
Q17 & Section 2 Architectural Context: The external systems, APIs, data sources, and actors are correctly derived from the repository content. & L \\
Q18 & Section 2.1 Use Case Diagram: The diagram correctly represents actors and use cases derived from the repository content. & L \\
Q19 & Section 3 Containers: The described containers, technologies, and communication protocols match the repository content. & L \\
Q20 &  Section 3.1 Component Diagram: The diagram correctly reflects the containers and externals as found in the repository. & L \\
Q21 & Section 4 Components: The identified components and interactions correspond to code-level relationships (imports, calls, dependencies). & L\\
Q22 &  Section 5 Code-Level: The entry points, key modules, and patterns correspond to actual code structure (no speculative elements). & L \\
Q23 & Section 5.1 Code-Level Diagram: The diagram includes only elements and relationships explicitly present in the repository & L \\
Q24 & Section 6 Cross-Cutting Concerns: The concerns listed (e.g., security, logging, configuration) are correctly represented. & L \\
Q25 & Section 7  Quality Attributes \& Rationale: The quality attributes and rationales are correctly represented (no speculative elements). & L \\
Q26 & Section 8 Deployment: The deployment information (nodes, environments) accurately reflects the repository’s infrastructure. & L \\
Q27 &  Section 8.1 Deployment Diagram: The diagram includes only infrastructure elements and connections declared in the repository. &  L \\
\midrule
\multicolumn{3}{l}{\textbf{Part 4 – Limitations and missing aspects (RQ4)}} \\
\midrule
Q28 &  Please describe any inconsistencies, inaccuracies, or unclear parts you noticed in the documentation. Which sections need improvement? & O \\
Q29 & What important architectural information is missing or insufficiently represented in the LLM-generated documentation? & O \\
Q30 & What improvements or changes would make the LLM-generated documentation more useful and trustworthy for your project? & O \\
\bottomrule
\end{tabularx}
\end{table*}

\noindent \textbf{Participants and Repositories.}
Participants were recruited through convenience sampling from the authors’ professional and academic networks. A total of $22$ developers agreed to take part in the study by contributing one of their software repositories and completing the survey after reviewing the automatically generated documentation. All participants had prior software development experience and were familiar with the repositories they submitted, ensuring knowledgeable and informed evaluations. The sample consists of $15$ researchers, $5$ PhD students, and $2$ software developers. Most respondents reported substantial development experience: $17$ indicated more than five years of practice, while only two reported one year or less. Participants also expressed strong familiarity with software architecture concepts, such as UML, architectural styles, and architectural patterns, with an average self-assessed score of $4.45/5$. Regarding their involvement with the analyzed repositories, $15$ participants stated that they had authored or maintained documentation (e.g., READMEs, architecture descriptions, or user guides). Their free-text responses referenced onboarding materials, installation guides, usage documentation, and high-level architectural overviews. Overall, the sample reflects practitioners and researchers with solid software engineering backgrounds and architectural knowledge. Many of them have direct responsibility for the documentation of the evaluated repositories.
Table~\ref{tab:repo-stats} summarizes the $22$ repositories contributed by participants, reporting for each project the main implementation languages, number of source files, and non-comment lines of code (LOC). The repositories exhibit substantial variability in size and technology stack: the number of files ranges from $3$ to $447$, and LOC from $81$ to $238{,}951$. The sample includes code bases predominantly written in Python, Java, C/C++, or JavaScript, often complemented by configuration and infrastructure artifacts such as YAML, JSON, XML, Dockerfiles, and shell scripts. This diversity indicates that the generated documentation was evaluated on both small and relatively large, multi-language repositories.

\noindent \textbf{Survey Execution and Data Analysis.}
Each participant provided a software repository of their choice, which was processed by our tool to generate the full architectural documentation. Participants were then asked to review the generated documentation and complete the questionnaire. No training phase was required, as the evaluation focused on the documentation itself rather than on the tool. Responses were collected individually.
For the closed-ended questions, we grouped answers according to the Likert-scale values and computed their frequencies to describe the overall distribution of participants’ perceptions. For the open-ended questions, we performed a thematic analysis~\cite{thematic_analysis} following standard qualitative research practices: two authors independently coded all textual answers, iteratively identifying and refining recurring themes related to strengths, weaknesses, and missing aspects of the generated documentation. Discrepancies between coders were discussed and resolved with the involvement of a third author, leading to a consolidated set of themes and ensuring consistency in the interpretation.

\section{Results}
\label{sec:Results}
In this section, we present the results of our empirical study by addressing each of the five research questions (RQ1–RQ5). For RQ1–RQ4, we report the quantitative findings from the questionnaire together with the qualitative insights derived from the open-ended responses, providing a comprehensive view of how developers evaluated the generated architectural documentation. For RQ5, we describe the time and cost required to automatically generate the documentation, characterizing the practicality of the proposed process.

\subsection{Answer to RQ1}
\begin{figure}[t]
  \centering
  \includegraphics[width=0.98\linewidth]{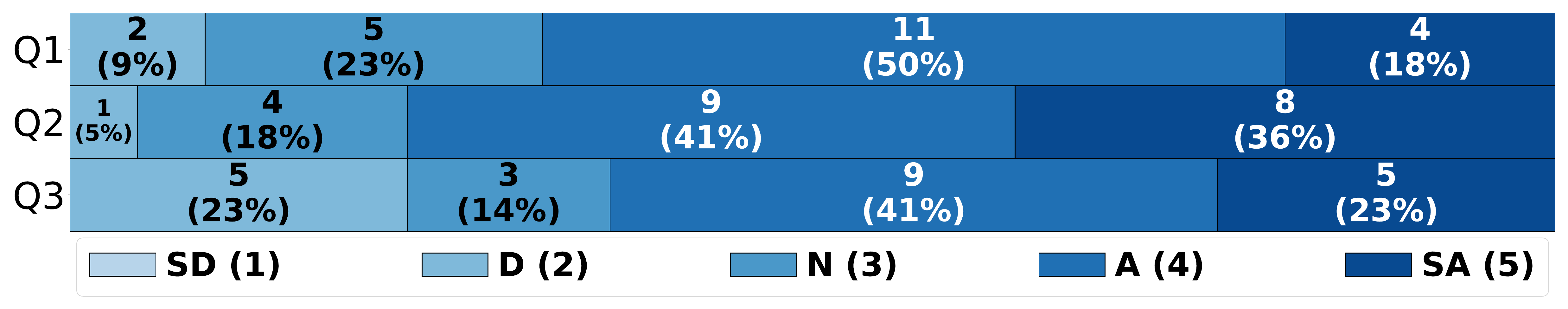}
  \caption{Distribution of participants’ Likert-scale ratings for RQ1}
  \label{fig:rq1-stacked}
\end{figure}

Figure~\ref{fig:rq1-stacked} shows the distribution of responses to Q1--Q4, indicating an overall positive perception of the LLM-generated architectural documentation. 
Q1, which probes whether the documentation provides valuable architectural insights, received predominantly positive ratings, with \textbf{A(11)} and \textbf{SA(4)}, while a smaller group selected \textbf{N(5)} or \textbf{D(2)}. 
A similar tendency appears in Q2, where respondents again expressed strong agreement, with \textbf{A(9)} and \textbf{SA(8)} accompanied by limited neutrality \textbf{N(4)} or disagreement \textbf{D(1)}. This suggests that the documentation generally supports comprehension of the system’s architectural structure. 
Perceptions of maturity (Q3) were more varied yet remained broadly positive, with \textbf{A(9)} and \textbf{SA(5)}. A subset of respondents selected \textbf{N(3)} or \textbf{D(5)}, indicating that some additional refinement or domain-specific adjustments may be needed before full adoption.

The open-ended responses further contextualize these findings by showing which sections developers found most valuable. The most recurrent theme was the usefulness of \textit{diagrams} (13 mentions), which were appreciated for making structural relationships easier to understand, as reflected in comments such as \textit{``Diagrams […] visually clarify complex information''} and \textit{``Class Diagrams […] make it easier to navigate dependencies''}. 

A second prominent theme relates to the value attributed to the \textit{Components} section (9 mentions), which several participants described as essential for understanding the system’s internal organization and for supporting modification tasks. As noted in the responses, \textit{``When I need to find which components I need to modify, I would check it here first''} and \textit{``[It] highlighted aspects of the codebase that I was not aware of.''}
Other sections were mentioned less frequently but still appreciated. The \textit{Deployment} (3 mentions) was valued for clarifying runtime and execution contexts, as reflected in \textit{``The deployment representation clearly identified dependencies''}. Similarly, \textit{Cross-Cutting Concerns} and \textit{Quality Attributes} (4 mentions) were recognized for surfacing system-wide behaviors and architectural considerations, with comments such as \textit{``Cross-Cutting and Quality Attributes […] the list of APIs''}.

Based on this evidence, the perceived value of the generated documentation can be summarized as follows:
\begin{tcolorbox}[colback=gray!10,boxrule=0.5pt,title=RQ1 Answer,boxsep=1pt,left=1pt,right=1pt,top=1pt,bottom=1pt]
Developers generally evaluated the LLM-generated documentation positively, finding it valuable for understanding system structure and, in many cases, mature enough for integration into their projects. Architectural diagrams, code-level details, and component views emerged as the most appreciated elements.
\end{tcolorbox}

\subsection{Answer to RQ2}

\begin{figure}[t]
  \centering
  \includegraphics[width=0.98\linewidth]{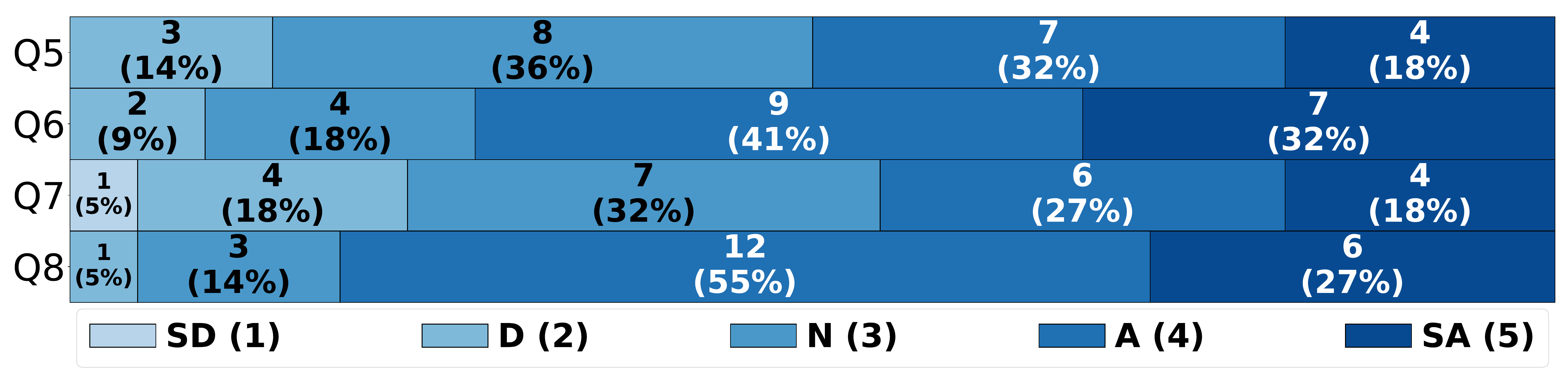}
  \caption{Distribution of participants’ Likert-scale ratings for RQ2}
  \label{fig:rq2-stacked}
\end{figure}

Figure~\ref{fig:rq2-stacked} shows that the overall perception of comprehensibility is positive. Clarity and organization (Q5) received predominantly favorable evaluations, with \textbf{A(7)} and \textbf{SA(4)}, while \textbf{N(8)} and a small number of negative ratings \textbf{D(3)} indicate that some parts of the documentation required additional effort to follow. 
Terminology (Q6) was viewed very positively, with \textbf{A(9)} and \textbf{SA(7)}, and only limited \textbf{N(4)} or \textbf{D(2)}. These results suggest that the architectural vocabulary used across the various sections, ranging from containers and components to code-level elements, was generally considered appropriate and consistent with standard architectural practice.

Opinions on redundancy (Q7) were more evenly distributed. Several respondents noted the presence of repetitive content \textbf{A(6)}, \textbf{SA(4)}, while many remained neutral \textbf{N(7)}, and fewer expressed disagreement \textbf{D(4)}, \textbf{SD(1)}. Overall, redundancy appears present but not sufficiently prominent to hinder comprehension.
With respect to the level of detail (Q8), participants expressed  positive perceptions, with \textbf{A(12)} and \textbf{SA(6)}, and only limited \textbf{N(3)} or \textbf{D(1)}. The documentation was considered sufficiently detailed to support understanding of architectural relationships, responsibilities, and interactions across the different abstraction levels represented in the template.

The open-ended responses highlight which parts of the documentation participants perceived as most comprehensible. The most frequently mentioned theme is the \textit{overall clarity of the documentation} (4 mentions), with respondents noting that \textit{``comprehensibility is good with all parts''} and \textit{``the entire documentation is easily readable and understandable''}. A second recurrent theme concerns the \textit{clarity of high-level architectural sections} (3 mentions), where system-level descriptions such as the Architectural Context and Containers were described as particularly easy to follow, as illustrated by comments like \textit{``Section 2 was the most comprehensible''} and \textit{``Section 3 is the most comprehensible […] provides valuable information''}.
Among less frequent themes, some respondents pointed to \textit{Cross-Cutting Concerns and Quality Attributes} (2 mentions) as especially clear, for example \textit{``Cross-Cutting Concerns''} and \textit{``Quality Attributes and Rationale''.} Finally, \textit{module-to-code and function mappings} (1 mention) were also singled out as understandable, as in \textit{``Module-to-Code Mapping and Key Functions''.} Overall, these observations indicate that participants generally perceived the documentation as comprehensible, particularly in its high-level narrative sections and in those parts that provide explicit structural mappings.
Overall, these quantitative patterns and qualitative remarks can be summarized as follows:

\begin{tcolorbox}[colback=gray!10,boxrule=0.5pt,title=RQ2 Answer,boxsep=1pt,left=1pt,right=1pt,top=1pt,bottom=1pt]
Developers generally find the documentation comprehensible: clarity, organization, terminology, and level of detail all received predominantly positive ratings, with only limited disagreement. Open-ended feedback reinforces this view, highlighting the overall readability of the documentation and, in particular, the clarity of high-level architectural sections.
\end{tcolorbox}

\subsection{Answer to RQ3}

\begin{figure}[h]
  \centering
  \includegraphics[width=0.98\linewidth]{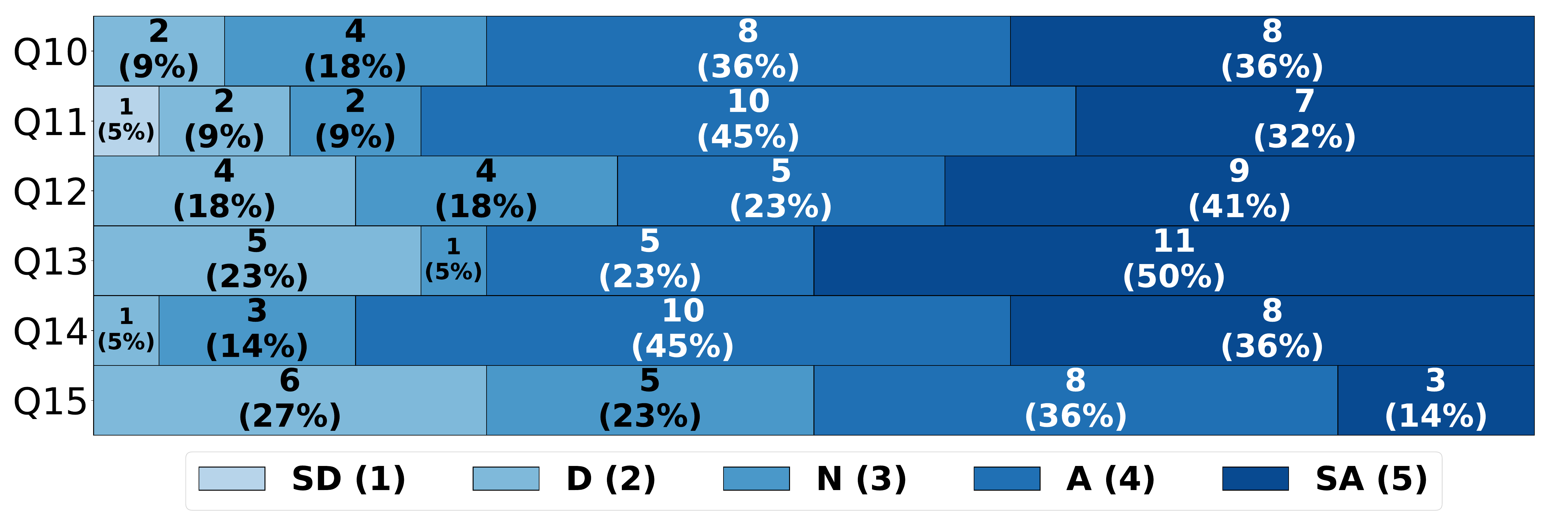}
  \caption{Distribution of participants’ Likert-scale ratings for RQ3 - General Part}
  \label{fig:rq3_plot}
\end{figure}

The responses to Q10--Q15, summarized in Figure~\ref{fig:rq3_plot}, assess several dimensions of accuracy, including architectural correspondence, internal consistency, completeness, correctness of relationships and responsibilities, and perceived reliability. Architectural correspondence (Q10) received largely positive ratings, with \textbf{A(8)} and \textbf{SA(8)}, while smaller groups selected \textbf{N(4)} or \textbf{D(2)}. Internal consistency (Q11) showed a similar trend, with \textbf{A(10)} and \textbf{SA(7)}, and only limited \textbf{N(2)}, \textbf{D(2)}, or \textbf{SD(1)}. Completeness (Q12) also followed this pattern, with \textbf{SA(9)} and \textbf{A(5)}, whereas \textbf{N(4)} and \textbf{D(4)} indicate that some omissions were noticed but did not dominate evaluations. Items Q13--Q15 confirm this trend: correctness of relationships (Q13) and responsibilities (Q14) was generally recognized, and perceived reliability (Q15), supported by \textbf{A(8)} and \textbf{SA(3)}, suggests that many participants viewed the documentation as a usable architectural reference.

\begin{figure}[h]
  \centering
  \includegraphics[width=0.98\linewidth]{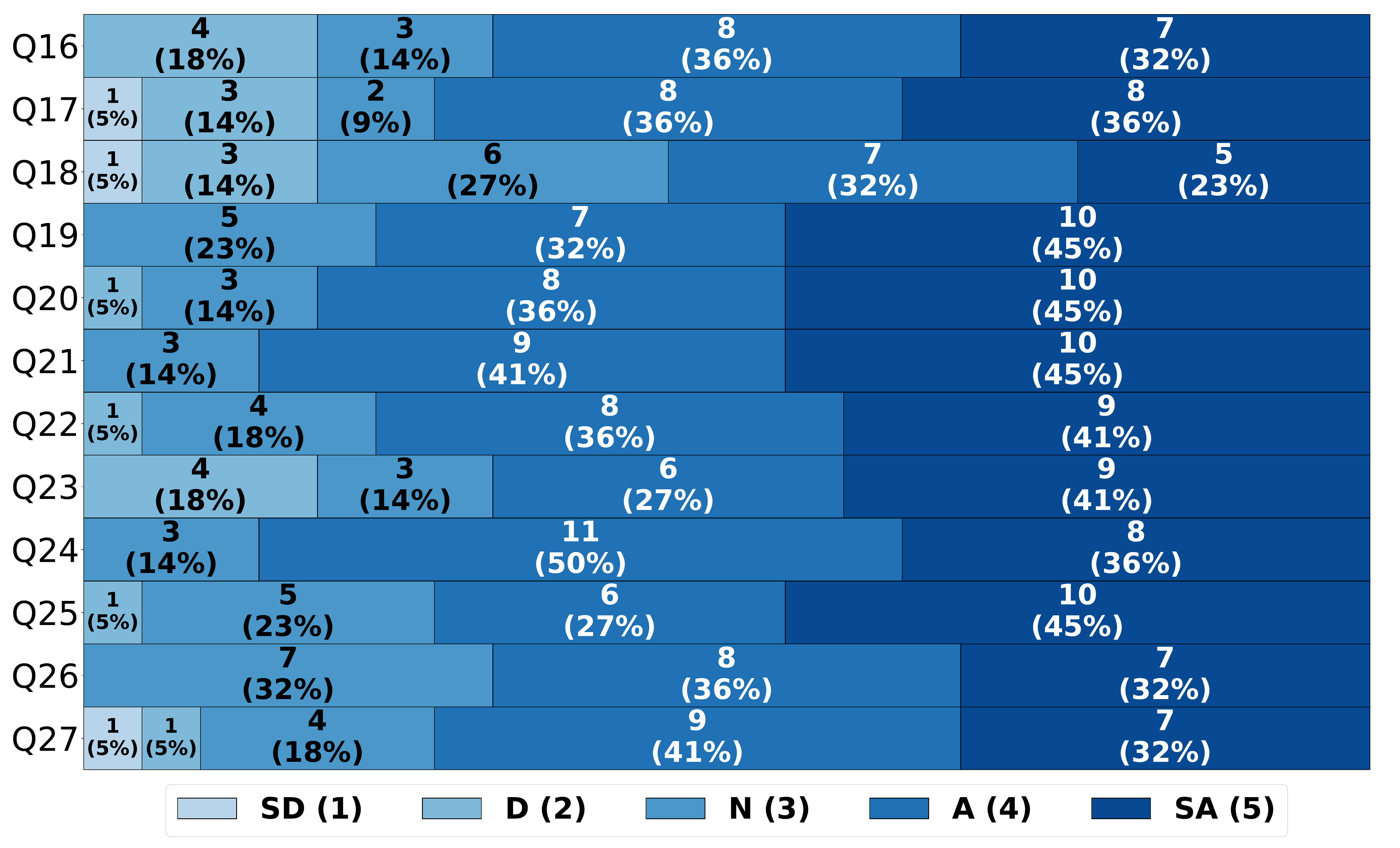}
  \caption{Distribution of participants’ Likert-scale ratings for RQ3 - Section Part}
  \label{fig:rq3_section}
\end{figure}

A more detailed view of section-level accuracy is provided by Q16--Q27 and summarized in Figure~\ref{fig:rq3_section}. Sections closely grounded in observable code evidence received the strongest agreement. The \textit{Components} section (Q21), describing the system’s internal logical structure, achieved the highest positive ratings with \textbf{SA(10)} and \textbf{A(9)}. \textit{Cross-Cutting Concerns} (Q24) showed a similarly strong pattern, with \textbf{SA(8)} and \textbf{A(11)}. High levels of agreement also emerged for the \textit{Component Diagram} (Q20) and the \textit{Containers} section (Q19) (18 and 17 positive responses, respectively). The \textit{Code-Level} section (Q22), capturing the concrete implementation of architectural elements in the source code, was likewise positively assessed, with 17 participants selecting \textbf{A} or \textbf{SA}. Collectively, these results indicate that sections reflecting architectural structure or explicit codebase artifacts were consistently perceived as accurate, whereas a smaller set of sections received more mixed evaluations. The \textit{System Overview} (Q16) and \textit{Architectural Context} (Q17) accumulated the highest numbers of negative ratings, \textbf{D(4)} and \textbf{D(3)}, suggesting that high-level descriptions were more prone to inaccuracies or omissions. 
The \textit{Use Case Diagram} (Q18) also showed weaker agreement, with several negative responses (\textbf{D(3)} and \textbf{SD(1)}). 
Additional structural diagrams also drew some negative feedback: the \textit{Code-Level Diagram} (Q23) received \textbf{D(4)}, while the \textit{Deployment Diagram} (Q27) received \textbf{D(1)} and \textbf{SD(1)}, indicating occasional mismatches with the underlying implementation or configuration.

Based on this combination of high-level and section-level assessments, the answer to RQ3 can be expressed as follows:
\begin{tcolorbox}[colback=gray!10,boxrule=0.5pt,title=RQ3 Answer,boxsep=1pt,left=1pt,right=1pt,top=1pt,bottom=1pt]
Developers perceived the LLM-generated documentation as accurate, indicating that it reflects the implemented architecture, is consistent across sections, and captures key structural elements. Accuracy was rated highest for code-related sections (Components, Containers, Code-Level), while more interpretive views (System Overview, Architectural Context, Use Case) received more mixed feedback. 
\end{tcolorbox}

\subsection{Answer to RQ4} 
Coding of the open-ended answers identified eight limitation categories, summarized in Table~\ref{tab:limitation_categories}. Beyond Q28--Q30, we also reviewed the open-ended responses from the other RQs to capture additional limitations or missing aspects. The most prominent category was \textit{Diagram Errors} (32 occurrences).

\begin{table}[b] 
\centering 
\small 
\caption{Limitation categories from open-ended responses.} \label{tab:limitation_categories} 
\begin{tabular}{l c} 
\hline \textbf{Category} & \textbf{Occurrences} \\ \hline Diagram errors & 32 \\ Deployment issues & 14 \\ Improvement suggestions & 13 \\ Missing information & 8 \\ Inconsistencies & 6 \\ Redundancy & 1 \\ \hline \end{tabular} 
\end{table} 

Participants frequently noted incomplete or misleading diagrams, mentioning for example \textit{`the class diagram is truncated and omits some of the classes''}, \textit{all the diagrams should be reviewed, and should be better presented […] it always seems a little too artificial’'}, and that some diagrams \textit{do not provide correct and useful information''}. These remarks reinforce that diagrammatic views were perceived as the most fragile artifacts. A second major theme relates to deployment issues (14 occurrences), particularly in runtime and infrastructure views. Reported problems included unclear relationships, such as \textit{`Deployment Diagram […] the relation between components is not clear''}, as well as incorrect or confusing arrows, for example \textit{Deployment (Local Workstation) has few arrows that should not exist''}. Other responses indicated that the deployment view should \textit{specify not only the production environment but also include an artifact repository to manage dependencies''} or that \textit{more details on the Deployment Diagram would be useful''}. Less frequent but still relevant themes include \textit{Inconsistencies} across sections (6 occurrences) and \textit{Missing Information} (8 occurrences). Some participants observed that \textit{`the overall components are captured, however […] wrapping of some of them is not well captured''} or noted that the LLM \textit{needs to be guided to avoid hallucinations and inconsistencies''}. Missing content was also highlighted, such as \textit{the class diagram is truncated and omits some of the classes''}, \textit{it misses the use case tables''}, or that respondents \textit{lack an in-depth description of this component''}. Finally, several responses offered general \textit{Improvement Suggestions} (13 occurrences), including \textit{`double-check with existing documentation if present''}, \textit{human inputs would surely enhance the quality of produced documentation''}, and requests for a shorter and better structured output to \textit{help users navigate it more easily''}. Redundancy was mentioned only once, with one participant noting that \textit{all sections are complete, but some content is repetitive''}.

In summary, the themes emerging from the open-ended responses converge toward the following answer to RQ4:

\begin{tcolorbox}[colback=gray!10,boxrule=0.5pt,title=RQ4 Answer,boxsep=1pt,left=1pt,right=1pt,top=1pt,bottom=1pt] 
Developers primarily point to problems in the diagrams, which are often described as incomplete, unclear, or inaccurate. Deployment views are the second most common source of issues, with respondents noting missing details or confusing relationships. Other limitations include occasional inconsistencies and missing information, along with suggestions to improve structure, reduce verbosity, and complement the generated output with human review. 
\end{tcolorbox}

\subsection{Answer to RQ5}
Table~\ref{tab:cost} reports the minimum, maximum, and average values of execution time and monetary cost for generating architectural documentation across the 22 analyzed repositories. Generation times ranged from 1m 50s to 4m 25.85s, with an average of approximately 3m 1s, showing that a complete architectural description can be produced in only a few minutes. Monetary costs ranged from \$0.35 to \$2.48, with an average of \$1.19 per repository, showing that an entire architecture document can be generated for roughly one to two dollars in API usage. These results suggest that the process is both time-efficient and economically lightweight compared to manual documentation effort.
\begin{table}[h]
\centering
\small
\caption{Costs of Generating the Documentation}
\label{tab:cost}
\begin{tabular}{lccc}
\toprule
\textbf{Metric} & \textbf{Min} & \textbf{Max} & \textbf{Mean} \\
\midrule
Generation Time & 1m 50s & 4m 25.85s & 3m 0.90s \\
Financial API Cost & \$0.35 & \$2.48 & \$1.19 \\
\bottomrule
\end{tabular}
\end{table}

Overall, these measurements of time and monetary cost can be summarized in the following answer:

\begin{tcolorbox}[colback=gray!10,boxrule=0.5pt,title=RQ5 Answer,boxsep=1pt,left=1pt,right=1pt,top=1pt,bottom=1pt]
Generating complete architectural documentation for a repository takes about three minutes on average, at an average API cost of \$1.19, making the process time-efficient and cost-effective compared to manual authoring.
\end{tcolorbox}

\subsection{Additional Practitioner Interest}

Beyond the formal evaluation, three participants expressed interest in using \textsc{CIAO} to document internal closed-source projects in their organizations, suggesting perceived practical value beyond the study setting. For example, one participant working on a safety-critical railway software component subject to strict regulations requested \textsc{CIAO}-generated documentation to support compliance activities, where architectural descriptions must be explicitly traceable to the source code. The participant found the structured representation of containers, components, and their dependencies particularly useful for SDLC activities such as test planning and execution.

Another participant, responsible for an internal tool used for penetration testing of edge–cloud infrastructures, faced challenges due to significant misalignment between the system’s architecture and its implementation, an instance of \textit{architectural drift}. Since the codebase had evolved faster than the documentation, \textsc{CIAO} was used to reconstruct an updated system-level description and restore architectural clarity.

A third participant, a software developer working in a company that provides digital services for public administrations, highlighted the absence of structured architectural documentation in several internal systems. They employed \textsc{CIAO} to generate system-level descriptions aimed at facilitating onboarding and improving knowledge transfer within their development team.

Taken together, these accounts illustrate early practitioner interest in deploying \textsc{CIAO} beyond open-source settings, particularly in contexts where regulatory compliance, architectural alignment, or team onboarding make structured system-level documentation especially valuable.

\section{Threats to Validity}
\label{sec:threats}

In this section, we discuss the main threats to validity of our study and the mitigation measures adopted to address them.

\noindent \textbf{Internal Validity.}
A potential threat lies in the subjective nature of the evaluation. Participants assessed documentation generated for repositories they had personally developed or maintained. While this familiarity reduces the risk of misinterpreting the underlying architecture, it may also introduce positive or negative bias in judging accuracy, completeness, or usefulness. To mitigate this threat, the questionnaire combined Likert-scale items with open-ended questions that asked participants to justify their ratings through concrete observations, thus reducing reliance on overall impressions.

\noindent \textbf{Construct Validity.}
This study evaluates developers’ perceptions of value, clarity, and accuracy rather than objective architectural correctness. The results also depend on the prompting strategy and the documentation template adopted in \textsc{CIAO}. An additional threat concerns the template itself, which was iteratively designed with a limited number of experts and may therefore cover only a restricted range of architectural concerns or shape how they are represented. To mitigate this, the questionnaire explicitly asked participants to assess the clarity, completeness, and usefulness of each section, providing an indirect validation of the template’s suitability for architecture-level analysis.

\noindent \textbf{External Validity.}
The study involved 22 participants, mostly from academic settings, and a diverse but not exhaustive set of repositories. This limits the generalizability of the findings to other developer populations, software domains, and LLM configurations. To partly mitigate this threat, we included repositories varying in size, language, and domain, although broader studies, especially with industrial participants, are needed to further assess the applicability of the results.

\noindent \textbf{Conclusion Validity.}
The thematic analysis of the open-ended responses may introduce interpretive bias. To mitigate this threat, two authors independently coded all responses and resolved disagreements with a third author, following established qualitative analysis practices to improve consistency and reliability in theme identification.

\section{Conclusions and Future Work}
\label{sec:conclusion}

This paper introduced \textsc{CIAO}, a structured process for automatically generating system-level architectural documentation from complete GitHub repositories using LLMs. \textsc{CIAO} combines a standards-oriented template with an LLM-based workflow grounded in ISO/IEC/IEEE~42010, SEI's \textit{Views \& Beyond}, and the C4 model.

Our empirical study with 22 developers suggests that the generated documentation is generally perceived as useful, understandable, and consistent with the architecture of the analyzed systems. Participants particularly appreciated the narrative sections and views closely linked to source code, while the main limitations concerned diagrammatic and deployment-related artifacts. The qualitative analysis also indicates that most inaccuracies are localized rather than structural, making the documentation a potentially useful architectural reference. In addition, the results show that the process is practically feasible, with manageable generation time and cost when using \texttt{GPT-5}.

Future work will focus on improving the reliability of diagrammatic views, which emerged as the weakest aspect of the generated documentation. Possible directions include integrating static or dynamic analysis and retrieval mechanisms tailored to architectural artifacts. We also plan to extend the evaluation to more system types, larger repositories, industrial contexts, and broader groups of developers. Finally, future studies may investigate how \textsc{CIAO} can support documentation and maintenance workflows, and whether human-in-the-loop refinement can further improve documentation quality and adoption.

\section*{Acknowledgment}
This work was supported by the Italian Ministry of Research, under the complementary actions to the NRRP “Fit4MedRob - Fit for Medical Robotics” Grant (\# PNC0000007).

\clearpage
\balance
\bibliographystyle{IEEEtran}
\bibliography{biblio}

\end{document}